\documentclass[aps,floatfix,preprint,showpacs,preprintnumbers,nofootinbib,superscriptaddress,natbib]{revtex4}

\usepackage{graphicx,float}
\usepackage{epsfig}		
\usepackage{dcolumn}
\usepackage{bm}
\usepackage{subfigure}

\usepackage[all]{xy}
\usepackage{amsmath,upgreek}
\usepackage{amssymb}

\usepackage{pdfpages}
\usepackage{color}
\usepackage{graphicx,epstopdf}
\usepackage[colorlinks,hyperindex]{hyperref}

\def\0{\mbox{\tiny $0$}}
\def\1{\mbox{\tiny $1$}}
\def\2{\mbox{\tiny $2$}}
\def\3{\mbox{\tiny $3$}}
\def\4{\mbox{\tiny $4$}}
\def\5{\mbox{\tiny $5$}}
\def\6{\mbox{\tiny $6$}}
\def\7{\mbox{\tiny $7$}}
\def\8{\mbox{\tiny $8$}}
\def\9{\mbox{\tiny $9$}}

\def\f14{\mbox{\tiny $\frac{1}{4}$}}

\def\bm#1{\mbox{\boldmath$#1$}}

\def\bb#1{\mbox{\footnotesize $(#1)$}}

\begin{document}

\title{Phase-space elementary information content of confined Dirac spinors}
\author{Alex E. Bernardini}
\email{alexeb@ufscar.br}
\affiliation{~Departamento de F\'{\i}sica, Universidade Federal de S\~ao Carlos, PO Box 676, 13565-905, S\~ao Carlos, SP, Brasil.}

\date{\today}

\begin{abstract}
Reporting about the Wigner formalism for describing Dirac spinor structures through a covariant phase-space formulation, the quantum information quantifiers for purity and mutual information involving spin-parity (discrete) and position-momentum (continuous) degrees of freedom are consistently obtained.
For Dirac spinor Wigner operators decomposed into Poincar\'e classes of $SU(2) \otimes SU(2)$ spinor couplings, a definitive expression for quantum purity is identified in a twofold way: firstly, in terms of phase-space positively defined quantities, and secondly, in terms of the {\em spin-parity traced-out} associated density matrix in the position coordinate representation, both derived from the original Lorentz covariant phase-space Wigner representation.
Naturally, such a structure supports the computation of relative (linear) entropies respectively associated to discrete (spin-parity) and continuous (position-momentum) degrees of freedom.
The obtained theoretical tools are used for quantifying (relative) purities, mutual information as well as, by means of the quantum concurrence quantifier, the spin-parity quantum entanglement, for a charged fermion trapped by a uniform magnetic field which, by the way, has the phase-space structure completely described in terms of Laguerre polynomials associated to the quantized Landau levels.
Our results can be read as the first step in the systematic computation of the elementary information content of Dirac-like systems exhibiting some kind of confining behavior.
\end{abstract}

\date{\today}
\maketitle

\section{Introduction}

Enlarging the access to elementary information features of physical systems without affecting the predictive power of quantum mechanics, the Wigner phase-space representation \cite{Wigner,Moyal,Case,Ballentine} of quantum mechanics has currently shed some light on the investigation of the frontiers between classical and quantum descriptions of Nature \cite{01A,02A,03A,JCAP18}.
Besides its ferramental pragmatic utility demanded by optical quantum mechanics \cite{Sch}, in the theoretical front, the Wigner quantum mechanics has also worked as a robust support for the non-commutative quantum mechanics \cite{Bastos3,Bastos3B,Bernardini13A,2015,Salomon,Bernardini13B,RSUP2,RSUP3,Bernardini13B2,PedroLeal}, for the description of the flux of quantum information in the phase-space \cite{Entro02,Steuernagel3,EPL18,Liouvillian} and, more generically, for probing quantumness and classicality for a relevant set of anharmonic quantum systems \cite{EPL18,PRA18,EJP19} as well as for quantitative modeling beyond the quantum physical framework \cite{Bert18}.

Recently, the framework has also been investigated in the context of lattice regularized quantum field theories and of lattice models of solid state physics \cite{Zub01,Zub02,Zub03}, from which a subtle connection with Dirac quantum mechanics is implied. 
In fact, the extension of the Wigner phase-space formalism to relativistic quantum mechanics, namely to the spinor space of Dirac equation solutions, was firmly stablished some decades ago \cite{1983,1986,1987}.
Gauge invariant relativistic quantum equations for fermionic Wigner operators were obtained from quantum electrodynamics and  expressed through a complete spinor decomposition procedure \cite{1983,1986,1987,Gao18}.

Motivated by such a Dirac-Wigner formalism for describing spinor structures -- through which a covariant phase-space derivation of Dirac equation solutions can be provided -- in the present work, the quantum information quantifiers for purity and mutual information involving spin-parity \cite{SU2,extfields,diraclike01} and position-momentum degrees of freedom shall be consistently obtained.
For Dirac spinor Wigner operators decomposed into Poincar\'e classes of $SU(2) \otimes SU(2)$ spinor couplings, a definitive expression for quantum purity shall be identified in a twofold way: one in terms of phase-space positively defined quantities and another one in terms of the {\em spin-parity traced-out} associated density matrix in the position coordinate representation.
Such equivalent results are derived from the original phase-space Wigner representation and, as expected, the complete structure that supports the computation of relative (linear) entropies respectively associated to discrete (spin-parity) and continuous (position-momentum) degrees of freedom shall also be identified.

Of course, this proposal involves the framework of quantum correlated states of spin and parity degrees of freedom written as the ground representation of the $SU(2) \otimes SU(2)$ symmetry\footnote{In this case, the $SU(2)$ is built as a subset of $SL(2, \mathbb{C})$, which is homomorphic to the homogeneous Lorentz group $SO(1,3)$.}, through which the Dirac spinor structures reflect the $SU(2) \otimes SU(2)$ spin-parity intrinsic properties of Dirac fermion free particle solutions. In this case, for free particle Dirac spinors in the coordinate space, the Hamiltonian equation expressed in natural units (i. e. with $c = \hbar = 1$) reads
\begin{equation}
\label{vvvv}
{H} \, \psi(\mbox{x}) = i \frac{\partial \, \psi(\mbox{x})}{\partial t} = (-i\mbox{\boldmath$\nabla$} \cdot {\mbox{\boldmath$\alpha$}} + m {\beta}) \,\psi(\mbox{x}) = (-i {\alpha}_i\partial^i + m {\beta}) \,\psi(\mbox{x}) = \pm E_p \,\psi(\mbox{x}),
\end{equation}
where $E_p = \sqrt {p^2 + m^2}$ are the eigenvalues and ``x'' is the short notation for the quadrivector $x_{\mu}$.
The so-called Dirac matrices (in their Dirac representation) -- the matrix operators ${\mbox{\boldmath$\alpha$}} =({\alpha}_x,\,{\alpha}_y,\,{\alpha}_z)$ and ${\beta}$ obeying the anticommuting relations, $
{\alpha}_i {\alpha}_j + {\alpha}_j {\alpha}_i = 2 \delta_{ij} {I}_4$, and ${\alpha}_i {\beta} + {\beta} {\alpha}_i =0$, for $i,j = x,y,z$, with $
{\beta}^2 = {I}_4$, where ${I}_N$ is the $N$-dim identity matrix -- can be decomposed into tensor products of Pauli matrices \cite{SU2}, ${\sigma}_i$, as ${\alpha}_i = {\sigma}_x^{(P)} \otimes {\sigma}_i^{(S)}$, for $i = x,y,z$ and ${\beta} = {\sigma}_z^{(P)} \otimes {I}_{2}^{(S)}$ such that one can recover ${H}$ in terms of Pauli matrix Kronecker products as
\begin{equation}
\label{erer}
{H} = \bm{p}\cdot ({\sigma}_x^{(P)} \otimes {\mbox{\boldmath$\sigma$}}^{(S)}) + m ( {\sigma}_z ^{(P)} \otimes {I}_{2}^{(S)}),
\end{equation}
with the superscripts $S$ and $P$ referring to degrees of freedom of spin and parity, in a picture that supports the interpretation of Dirac state vectors as double-doublets of the $SU(2)\otimes SU(2)$ \cite{SU2,diraclike01,diraclike02,PRA2018}.

Given that Dirac-like systems with the intrinsic information content supported by the above structure have been considered in the emulation of the relativistic quantum mechanics properties for several platforms mimicking low energy physics \cite{diraclike01,diraclike02,teleport}, the investigation of how to suit a localizing description of Dirac spinors in the phase-space representation of quantum mechanics is a pressing issue which give reasons to our interests in describing the elementary information content of localized Dirac spinors in the Wigner framework.

From an enlarged overview perspective, quantum information quantifiers for quantum entanglement and decoherence have already been tested through Dirac-like systems emulated, for instance, by trapped ion platforms adapted for detecting local quantum correlations \cite{Nat01,Nat03}, and for simulating open quantum systems and quantum phase transitions \cite{Nat02,Nat04,Nat05}, where a phenomenological access to manipulate quantum information properties of trapped ions has been provided \cite{n004,n005,n006}. In particular, the ionic Jaynes-Cummings Hamiltonian dynamics simulates a series of relativistic Dirac-like effects \cite{n001,n002,new01} which is mapped by a $\mbox{SU}(2)\otimes \mbox{SU}(2)$ bi-spinor structure typical from such ionic systems, and which engenders the entanglement of two-{\em qubit} states, in this case, related to Hilbert subspaces associated to total angular momentum and to its projection onto the direction of an external magnetic field which lifts the ion's internal levels \cite{diraclike01,diraclike02}.
Trapped ions are closed systems over which localizing interactions can generate decoherence and degradation of quantum correlations between the subsystems \cite{intronoise00, intronoise01, intronoise02}. Such systems devised to eliminate decoherence and dephasing generated by a global noise \cite{FreqEst} in qubit memory engineering \cite{NoiseTrap01, NoiseTrap02, NoiseTrap03, NoiseTrap04}. In such a context, for instance, local effects related to Dirac-like localized structures can be relevant in describing quantum decoherence, at least, from a preliminar theoretical perspective.

Still from a theoretical perspective, the $SU(2) \otimes SU(2)$ Dirac bispinor structure exhibits an entanglement profile that is invariant under Lorentz rotations, quantitatively preserved at any reference frame.
The information related to the spin polarization and the $SL(2,\mathbb{C})\otimes SL(2,\mathbb{C})$ correlated information related to the intrinsic parity are mutually involved into Dirac bispinor correlations.
Given that parity operations exchange two irreducible representatons ({\em irreps}) of the Poincar\'{e} group, an adequate procedure for discussing entanglement must be yielded in terms of {\em irreps} of the complete Lorentz group.
It is required by the condition that the $SU(2) \otimes SU(2)$ symmetry is only one of the inequivalent representations enclosed by the $SL(2,\mathbb{C})\otimes SL(2,\mathbb{C})$ symmetry.
Therefore, the inclusion of confining properties in that formalism, namely for obtaining the for spin-parity quantum concurrence as an entanglement quantifier is relevant along the discussion of quantum correlations for physical system which can be covered by the $SL(2,\mathbb{C})\otimes SL(2,\mathbb{C})$ symmetry.

Again, by identifying the map to low energy Dirac-like systems, the relation between the intrinsic entanglement of non-confined Dirac equation solutions and the entanglement of low energy mesoscopic quantum systems has already been investigated, for instance, for bilayer graphene excitations, in the stable configuration of the Bernal stacking \cite{graph03,graph04}. In this case, the tight binding Hamiltonian (including both bias voltage and mass terms) \cite{graph03,graph04,diraclike02,Predin}, when it is written in the reciprocal space, can be directly identified with a modified Dirac Hamiltonian including external pseudovector and tensor potentials, which maps lattice and layers associated Hilbert spaces for graphene \cite{diraclike02,Predin}.

As a first step in close connection with the above-mentioned quantum scenarios, the obtained theoretical tools are used for quantifying (relative) purities and the mutual information for a charged fermion trapped by a uniform magnetic field which, by the way, has the phase-space structure completely described in terms of Laguerre polynomials associated to the quantized energy Landau levels. In such a context, the issue of spin-parity quantum concurrence for measuring the quantum entanglement is also worked out.
Our work can be read as a preliminary systematic approach for computing the elementary information content of Dirac-like systems exhibiting some kind of confining behavior.

Our manuscript is thus organized as follows.
In Section II, the Wigner formalism for describing Dirac spinor structures \cite{1986,1987} is recovered in the context of a covariant phase-space formulation of Dirac equation solutions through which the Dirac spinor Wigner operators decomposed into Poincar\'e classes of $SU(2) \otimes SU(2)$ spinor couplings are identified.
In Section III, the quantum information quantifiers for purity and mutual information involving spin-parity and position-momentum degrees of freedom are obtained. In particular, the definitive expression for quantum purity is identified in terms of phase-space positively defined quantities as well as in terms of the {\em spin-parity traced-out} associated density matrixe in the position coordinate representation.
As a generic application, in Section IV, the (relative) purities and the mutual information content for a charged fermion trapped by a uniform magnetic field are analytically obtained, given that its phase-space structure can be completely described in terms of Laguerre polynomials associated to the Landau levels.
As mentioned above, the issue of spin-parity quantum entanglement is also addressed along this section.
Finally, our conclusions are drawn in Section V, in order to point to extensions of this formalism.

\section{Wigner-Dirac functionals and the spinor decomposition}

The quantum mechanical analogue of the classical phase-space distribution function is the so-called Wigner function,  $\omega(x,\, k_x)$ which is obtained through the Weyl transform of a generic quantum operator, $\hat{O}$,
\begin{equation}\small
O^W(x,\, k_x)
= 2\hspace{-.2cm} \int_{_{-\infty}}^{^{+\infty}} \hspace{-.5cm}du\,\exp{\left[2\,i \,k_x\, u/\hbar\right]}\,\langle x - u | \hat{O} | x + u \rangle=2\hspace{-.2cm} \int_{_{-\infty}}^{^{+\infty}} \hspace{-.5cm} dr \,\exp{\left[-2\, i \,x\, r/\hbar\right]}\,\langle k_x - r | \hat{O} | k_x + r\rangle,
\end{equation}\normalsize
when it is applied into a density matrix operator, $\hat{\rho} = |\psi \rangle \langle \psi |$, as to return
\begin{equation}
 h^{-1} \hat{\rho} \to  \omega(x, k_x) =  (\pi\hbar)^{-1} 
\int_{_{-\infty}}^{^{+\infty}} \hspace{-.5cm}du\,\exp{\left[2\, i \, k_x \,u/\hbar\right]}\,
\psi^{\ast}(x - u)\,\psi(x + u),
\end{equation}
which can also be interpreted as the Fourier transform of the off-diagonal elements of the associated density matrix.

By applying the Weyl's correspondence principle onto the relativistic field theory, the phase-space distribution function reads the ensemble average of the Wigner operator, which is written in the following covariant form \cite{1987},
\begin{eqnarray}
\hat{W}_{\xi\lambda}(\mbox{x},\, \mbox{k}) &=&  {\pi^{-4}}\int d^4{u} \,\exp{\left[- 2\, i \, k^{\mu}u_{\mu}\right]}\,
\overline{\psi}_{\lambda}(\mbox{x})\,\exp[i u^{\mu}\partial^{\dagger}_{\mu}]\,\exp[- i u^{\mu}\partial_{\mu}]\,\psi_{\xi}(\mbox{x})
\nonumber\\
\Rightarrow\,
 {W}_{\xi\lambda}(\mbox{x},\, \mbox{k}) &=&{\pi^{-4}}\int d^4{u} \,\exp{\left[- 2\, i \, k^{\mu}u_{\mu}\right]}\,
\overline{\psi}_{\lambda}(\mbox{x} -\mbox{u})\,\psi_{\xi}(\mbox{x} +\mbox{u}),
\label{344}
\end{eqnarray}
where $\hbar$ has been set equal to unity, and the exponentialized derivatives -- $\partial_{\mu}^{\dagger}$ (to the left) and $\partial_{\mu}$ (to the right) -- are the generators of $4$-dim translations and the Heisenberg spinor operators are identified by the subindices, $\lambda$ and $\xi$, which define the $4 \times 4$ components of the above Wigner operator.
With the $4$-dim momentum operator identified by $\hat{\mbox{p}} \equiv \hat{p}_{\mu} = (i/2)(\partial_{\mu} - \partial_{\mu}^{\dagger})$, the integration over $u$ results into 
\begin{equation}
W_{\xi\lambda}(\mbox{x},\, \mbox{k}) =\langle :\overline{\psi}_{\lambda}(\mbox{x})\,\delta^{(4)}(\mbox{k} - \hat{\mbox{p}}) \psi_{\xi}(\mbox{x}): \rangle
\end{equation}
where the colons indicate the normal ordering with respect to the vacuum state, and brackets indicate averaged values.

In particular, the trace of $W_{\xi\lambda}(\mbox{x},\, \mbox{k})$, i. e. $\sum_{\lambda} W_{\lambda\lambda}(\mbox{x},\, \mbox{k})$, returns the Lorentz scalar density of Dirac fermions at aspace-time point $x_{\mu}$, with quadrimomentum $k_{\mu}$. It can be noticed from the fermion vector current given by
\begin{equation}\small
\label{noen}e^{-1} \langle j_{\mu}(\mbox{x})\rangle = \langle :\overline{\psi}(\mbox{x})\,\gamma_{\mu}\,\psi(\mbox{x}): \rangle =
Tr\left[\int d^{4}k\,\gamma_{\mu}\,\langle: \hat{W}(\mbox{x},\, \mbox{k}):\rangle\right]
=Tr\left[\int d^{4}k\,\gamma_{\mu}\,{W}(\mbox{x},\, \mbox{k})\right],
\end{equation}\normalsize
which can work as the setup for obtaining the equation of motion for the Wigner function \cite{Gao18,1987}. However, unlike  classical and non-relativistic quantum distribution functions, the relativistic Wigner function depends on time-like coordinates, which naturally introduces some manipulation difficulties.
On one hand, in gauge quantum field theories, for instance, it demands for modifications in order to ensure local gauge invariance, given that the invariant translations defined by $ i y^{\mu}\partial_{\mu}$ are not well defined in gauge theories \cite{1987}. 
On the other hand, the physical observables related to averaged value calculations do not return finite values for the time-like component of the volume integration.
In this case, one should be concerned with the problems of the Wigner framework in providing an effective predictability in case of frame-defined calculations, which can only be circumvented when the localization properties are correctly addressed \footnote{Similar to what happens for equal time procedures in the computation of quantum propagators in canonical quantum field theories.}.

Fortunately, in the context of the Wigner formalism, considering the notation from Eq.~(\ref{344}), the most appealing configuration for the spinor fields that engenders the Wigner distribution is given in terms of
\begin{equation}
\psi_{\lambda}({ \mbox{x} \pm \mbox{u}}) =\psi_{\lambda}(\mathbf{x} \pm \mathbf{u}) \, \exp[-i k^{(\lambda)}_0(t + \tau)],
\end{equation}
where $\lambda$ is the spinor index, and $k_0$, $t$ and $\tau$ are the respective time-like components of $k_{\mu}$, $x_{\mu}$ and $u_{\mu}$.
It supports the definition of an auxiliary operator identified by the energy-averaged operator
\begin{eqnarray}\label{122223}
\omega_{\xi\lambda}(\mathbf{x},\,\mathbf{k};\,t) &=& 
\int_{_{-\infty}}^{^{+\infty}}\hspace{-.5 cm} d\mathcal{E}\, {W}_{\xi\lambda}(\mbox{x},\, \mbox{k})
\nonumber\\&=& \pi^{-1}\,\exp[i (k^{(\lambda)}_0 - k^{(\xi)}_0)t]\,\int_{_{-\infty}}^{^{+\infty}}\hspace{-.5 cm} d\tau
\int_{_{-\infty}}^{^{+\infty}}\hspace{-.5 cm} d\mathcal{E}\,\exp[-i (2\mathcal{E} - k^{(\lambda)}_0 - k^{(\xi)}_0)\tau]
\nonumber\\&&\qquad
\times \,\pi^{-3}\int d^{3}\mathbf{u}\,\exp[2i \mathbf{k}\cdot\mathbf{u}]\,
 \overline{\psi}_\lambda(\mathbf{x} -\mathbf{u})\, \psi_{\xi}(\mathbf{x} + \mathbf{u})
 \nonumber\\&=&
 2\,\exp[i (k^{(\lambda)}_0 - k^{(\xi)}_0)t]\,\int_{_{-\infty}}^{^{+\infty}}\hspace{-.5 cm} d\mathcal{E}\,\delta(2\mathcal{E} - k^{(\lambda)}_0 - k^{(\xi)}_0)
\nonumber\\&&\qquad
\times \,\pi^{-3}\int d^{3}\mathbf{u}\,\exp[2i \mathbf{k}\cdot\mathbf{u}]\,
 \overline{\psi}_\lambda(\mathbf{x} -\mathbf{u})\, \psi_{\xi}(\mathbf{x} + \mathbf{u})
 \nonumber\\&=&
 \pi^{-3}\exp[i (k^{(\lambda)}_0 - k^{(\xi)}_0)t]
 \int d^{3}\mathbf{u}\,\exp[2i \mathbf{k}\cdot\mathbf{u}]\,
 \overline{\psi}_\lambda(\mathbf{x} -\mathbf{u})\, \psi_{\xi}(\mathbf{x} + \mathbf{u}),\qquad
\end{eqnarray}
which is noway affected by time-like $\tau$ and $\mathcal{E}$ {\em useless} integrations, and which exhibits the properties of an Euclidian phase-space quasi-distribution of probabilities normalized by
\begin{equation}
 \pi^{-3}
 \int d^{3}\mathbf{x} \int d^{3}\mathbf{k}\,\omega_{\xi\lambda}(\mathbf{x},\,\mathbf{k};\,t) = \delta_{\xi\lambda},
\end{equation}
which, according to such an approach, also results into an {\em Euclidian version} of Eq.~(\ref{noen}),  
\begin{equation}
\label{noen2}e^{-1} \langle j^{\lambda\xi}_{\mu}(\mathbf{x};\,t)\rangle =Tr\left[\int d^{3}\mathbf{k}\,\gamma_{\mu}\,\omega_{\xi\lambda}(\mathbf{x},\,\mathbf{k};\,t)\right]
\end{equation}
which, of course, describes stationary states for $\lambda = \xi$ and gives normalized probability distributions as
\begin{equation}
e^{-1}  \int d^{3}\mathbf{x} \langle j_{0}(\mathbf{x};\,t)\rangle = \int d^{3}\mathbf{x}\, Tr\left[\int d^{3}\mathbf{k}\,\gamma_{0}\,\omega(\mathbf{x},\,\mathbf{k};\,t)\right] =1,
\end{equation}
where the spinor indices were momentarily suppressed.

Finally, in order to clear up the meaning of the spinor structure into the Wigner phase-space formalism, one can decompose the spinor structure of the Wigner function in the matrix form in terms of the 16 independent generators of the Clifford algebra. 
For the conventional basis in a Dirac representation form, with the gamma matrices written as $\gamma_{0} = \beta$, $\gamma_{j} = \beta\alpha_j$, and with $\{\gamma_{\mu},\gamma_{\nu}\} = 2g_{\mu\nu}$, $\{\gamma_{\mu},\gamma_{5}\} = 0$ and $\sigma_{\mu\nu} = (i/2)[\gamma_{\mu},\gamma_{\nu}]$, one has \cite{1987}
\begin{equation}\label{51}
\omega(\{q\}) \equiv
\mathcal{S}(\{q\})+
i\,\gamma_{5}\,{\Pi}(\{q\})+
\gamma_{\mu}\,\mathcal{V}^{\mu}(\{q\})+
\gamma_{\mu}\gamma_{5}\,\mathcal{A}^{\mu}(\{q\})+
\frac{1}{2}\sigma_{\mu\nu}\mathcal{T}^{\mu\nu}(\{q\}),
\end{equation}
with $\{q\}\equiv \{\mathbf{x},\,\mathbf{k};\,t\}$, and where the sum brings up successive scalar, pseudoscalar, vector, axial vector, and antisymmetric tensor contributions (under Lorentz transformation), respectively described by
\begin{eqnarray}
\omega(\{q\}) \equiv
\mathcal{S}(\{q\}) &=&\frac{1}{4}Tr[\omega(\{q\})],\\
{\Pi}(\{q\})&=&-\frac{i}{4}Tr[\gamma_{5}\,\omega(\{q\})],\\
\mathcal{V}^{\mu}(\{q\})&=&\frac{1}{4}Tr[\gamma^{\mu}\,\omega(\{q\})],\\
\mathcal{A}^{\mu}(\{q\})&=&\frac{1}{4}Tr[\gamma_{5}\gamma^{\mu}\,\omega(\{q\})],\\
\mathcal{T}^{\mu\nu}(\{q\})=-\mathcal{T}^{\nu\mu}(\{q\})&=&\frac{1}{4}Tr[\sigma^{\mu\nu}\,\omega(\{q\})].
\label{55}\end{eqnarray}

From now on, once the above introduced frame-defined approach has been stablished, one has the essential elements for computing quantum information quantifiers for phase-space Dirac-like Wigner quantum systems.  

\section{Quantum purity and mutual information under spatial localization}

Considering the statistical aspects related to the definition of the density matrix operators, where the Wigner formalism admits extensions from pure states to statistical mixtures, the phase-space purity $\mathcal{P}$ in $1$-dim is read as
\begin{equation}
\mathcal{P} = Tr_{\{x,k_x\}}[\hat{\rho}^2] = 2\pi\int^{+\infty}_{-\infty} \hspace{-.5cm} {dx}\int^{+\infty}_{-\infty} \hspace{-.5cm}{dk_x}\,W(x,\, k_x)^2,
\label{pureza}
\end{equation}
where the factor $2\pi$ is introduced to satisfy the constraint of $Tr_{\{q,p\}}[\hat{\rho}^2] = 1$ for pure states.

The straightforward extension of the above definition to Dirac spinors results into
\small\begin{eqnarray}
\mathcal{P}&=& (2\pi)^3\int\hspace{-.2cm}d^3\mathbf{x}\int\hspace{-.2cm}d^3\mathbf{k} \, Tr\left[\left(\gamma^{0}\omega(\mathbf{x},\,\mathbf{k};\,t)\right)^2\right] = 
(2\pi)^3\int\hspace{-.2cm}d^3\mathbf{x}\int\hspace{-.2cm}d^3\mathbf{k} \, Tr\left[\omega(\mathbf{x},\,\mathbf{k};\,t)\,\omega^{\dagger}(\mathbf{x},\,\mathbf{k};\,t)\right]
\end{eqnarray}\normalsize
which, by the way, results into a positive definite quantity given by (cf. Eqs.~(\ref{51})-(\ref{55}))
\begin{eqnarray}
\mathcal{P}&=& 4\times(2\pi)^3\int\hspace{-.2cm}d^3\mathbf{x}\int\hspace{-.2cm}d^3\mathbf{k} 
\left[\mathcal{S}^2 + \Pi^2 + \mathcal{V}_{\mu}\mathcal{V}^{\tilde{\mu}} + \mathcal{A}_{\mu}\mathcal{A}^{\tilde{\mu}} + \frac{1}{2}\mathcal{T}_{\mu\nu}\mathcal{T}^{\tilde{\mu}\tilde{\nu}}\right],
\end{eqnarray}
where the {\em tilde} ``$^{\sim}$'' over the indices ``$\mu$'' and ``$\nu$'' denotes the conversion of Minkowskian products into auxiliary Euclidian products defined by 
$$\mathcal{V}_{\mu}\mathcal{V}^{\tilde{\mu}} = \sum_{s=0}^3 \mathcal{V}_s^2,\qquad \mathcal{A}_{\mu}\mathcal{A}^{\tilde{\mu}} = \sum_{s=0}^3 \mathcal{A}_s^2,\qquad \mbox{and} \qquad \mathcal{T}_{\mu\nu}\mathcal{T}^{\tilde{\mu}\tilde{\nu}} = \sum_{s=0}^3 (\mathcal{T}_{0s}^2 +\sum_{j=1}^3\mathcal{T}_{js}^2).$$

An equivalent alternative (also $3$-dim reduced) expression for $\mathcal{P}$ can be obtained directly from Eq.~(\ref{122223}) as
\begin{eqnarray}
\mathcal{P}&=& (2\pi)^3\int\hspace{-.2cm}d^3\mathbf{x}\int\hspace{-.2cm}d^3\mathbf{k}\int \frac{d^3\mathbf{u}}{\pi^3}\int \frac{d^3\mathbf{w}}{\pi^3} 
\exp[2i\,\mathbf{k}\cdot(\mathbf{u}-\mathbf{w})]
\nonumber\\
&& \qquad\times Tr_{\xi\xi'}\left[\gamma_{0} \overline{\psi}_{\lambda}(\mathbf{x}-\mathbf{u}){\psi}_{\xi}(\mathbf{x}+\mathbf{u})\,\gamma_{0}\,\overline{\psi}_{\lambda}(\mathbf{x}-\mathbf{w}){\psi}_{\xi'}(\mathbf{x}+\mathbf{w}) \right]\nonumber\\
&=& 2^3(2\pi)^3 \pi^{-3}\int\hspace{-.2cm}d^3\mathbf{x}\int\hspace{-.2cm}d^3\mathbf{w}\int\hspace{-.2cm}d^3\mathbf{w} \,\,\delta^{(3)}(2(\mathbf{u}-\mathbf{w}))\nonumber\\
&& \qquad\times \,\,Tr_{\xi\xi'}\left[\gamma_{0} \overline{\psi}_{\lambda}(\mathbf{x}-\mathbf{u}){\psi}_{\xi}(\mathbf{x}+\mathbf{u})\,\gamma_{0}\,\overline{\psi}_{\lambda}(\mathbf{x}-\mathbf{u}){\psi}_{\xi'}(\mathbf{x}+\mathbf{u}) \right]\nonumber\\
&=& 2^3\int\hspace{-.2cm}d^3\mathbf{x}\int\hspace{-.2cm}d^3\mathbf{u}\, Tr_{\xi\xi'}\left[\gamma_{0} \overline{\psi}_{\lambda}(\mathbf{x}-\mathbf{u}){\psi}_{\xi}(\mathbf{x}+\mathbf{u})\,\gamma_{0}\overline{\psi}_{\lambda}(\mathbf{x}-\mathbf{u}){\psi}_{\xi'}(\mathbf{x}+\mathbf{u}) \right],
\end{eqnarray}
where , again, the factor $(2\pi)^3$ was introduced so as to satisfy the unitary constraint for the purity.
By noticing that $\gamma_{0}\,\omega\,\gamma_{0} = \omega^{\dagger}$, and consequently,
\begin{eqnarray}
\lefteqn{Tr_{\xi\xi'}\left[\gamma_{0} \overline{\psi}_{\lambda}(\mathbf{x}-\mathbf{u}){\psi}_{\xi}(\mathbf{x}+\mathbf{u})\,\gamma_{0}\overline{\psi}_{\lambda}(\mathbf{x}-\mathbf{u}){\psi}_{\xi'}(\mathbf{x}+\mathbf{u}) \right]}\nonumber\\
&=& Tr_{\xi\xi'}\left[\overline{\psi}_{\xi}(\mathbf{x}+\mathbf{u})\left({\psi}_{\lambda}(\mathbf{x}-\mathbf{u})\,\overline{\psi}_{\lambda}(\mathbf{x}-\mathbf{u})\right){\psi}_{\xi'}(\mathbf{x}+\mathbf{u})\right]
\nonumber\\
&=& \varrho(\mathbf{x}-\mathbf{u})\varrho(\mathbf{x}+\mathbf{u}),
\end{eqnarray}
with $$\varrho(\mathbf{r}) = {\psi}_{\lambda}(\mathbf{r})\,\overline{\psi}_{\lambda}(\mathbf{r})= \sum_{\lambda=1}^4\overline{\psi}_{\lambda}(\mathbf{r})\,{\psi}_{\lambda}(\mathbf{r}) = \overline{\psi}(\mathbf{r})\,{\psi}(\mathbf{r}),$$
one obtains
\begin{eqnarray}
\mathcal{P}&=& \int\hspace{-.2cm}d^3\mathbf{x}\int\hspace{-.2cm}d^3\mathbf{u}\,\varrho(\mathbf{x}-\mathbf{u}/2)\varrho(\mathbf{x}+\mathbf{u}/2), \end{eqnarray}
which allows for obtaining the purity directly from the coordinate representation of the Dirac spinors. 

One can go further in order to compute the mutual information involving spin-parity and position-momentum degrees of freedom.
In this case, the partial trace related to spin-parity degrees of freedom is identified by
$\rho(\mathbf{x},\,\mathbf{k};\,t) = Tr_{\xi\lambda}[\omega_{\xi\lambda}(\mathbf{x},\,\mathbf{k};\,t)\,\gamma_0]$ as well as the partial trace related to position-momentum degrees of freedom is identified by the averaged integration of $\omega_{\xi\lambda}\,\left(\gamma_0\right)$ over the phase-space volume $V$, with $dV = d^3\mathbf{x}\,d^3\mathbf{k}$,
$$\langle\omega_{\xi\lambda}\,\gamma_0\rangle = \langle\omega_{\xi\lambda}\rangle \gamma_0= \int\hspace{-.2cm}d^3\mathbf{x}\int\hspace{-.2cm}d^3\mathbf{k}\,\,\omega_{\xi\lambda}(\mathbf{x},\,\mathbf{k};\,t) \gamma_0.$$

Then, the relative linear entropies related to discrete spin-parity ($SP$) and continuous position-momentum $(\mathbf{x},\,\mathbf{k})$ variables result, respectively, into
\begin{eqnarray}
\mathcal{I}^{SP} &=& 1 - Tr\left[\left(\langle\omega_{\xi\lambda}\rangle \gamma_0\right)^2\right]
\nonumber\\&=&1- 4\left[\langle\mathcal{S}\rangle^2 + \langle\Pi\rangle^2 + \langle\mathcal{V}_{\mu}\rangle\langle\mathcal{V}^{\tilde{\mu}}\rangle + \langle\mathcal{A}_{\mu}\rangle\langle\mathcal{A}^{\tilde{\mu}}\rangle + \frac{1}{2}\langle \mathcal{T}_{\mu\nu}\rangle\langle \mathcal{T}^{\tilde{\mu}\tilde{\nu}}\rangle\right]
,\end{eqnarray}
and
\begin{eqnarray}
\mathcal{I}_{\{\mathbf{x},\,\mathbf{k}\}} &=& 1 - (2\pi)^3 
\int\hspace{-.2cm}d^3\mathbf{x}\int\hspace{-.2cm}d^3\mathbf{k}\,\left(\rho(\mathbf{x},\,\mathbf{k};\,t)\right)^2\nonumber\\&=&
1 - 16 \times (2\pi)^3 \langle \mathcal{V}_0^2\rangle.
\end{eqnarray}

Of course, the above identified elementary structures are much more appealing when applied into the investigation of complex Dirac-like quantum problems, as for instance, those emulated by trapped ion platforms \cite{diraclike01} and bilayer graphene configurations \cite{diraclike02}.  
As reported in Refs.~\cite{extfields,SU2}, the spin-parity intrinsic structure of Dirac spinors admits the inclusion of  interacting properties as they appear in Dirac Hamiltonians like \cite{diraclike01,diraclike02}
\begin{eqnarray}
\label{E04T}
{H}  &=& Q\,A^0\bb{\bm{x}}\,{I}_4+ {\beta}( m + \phi_S \bb{\bm{x}} ) + {\mbox{\boldmath$\alpha$}} \cdot ({\bm{p}} - Q\bm{A}\bb{\bm{x}}) + i {\beta} {\gamma}_5 m\bb{\bm{x}} - {\gamma}_5 q\bb{\bm{x}} + {\gamma}_5 {\mbox{\boldmath$\alpha$}}\cdot\bm{W}\bb{\bm{x}} \nonumber \\
&+& i {\bm{\gamma}} \cdot [ \zeta_a \bm{B}\bb{\bm{x}} + \kappa_a\, \bm{E}\bb{\bm{x}}  \,] + {\gamma}_5 {\bm{\gamma}}\cdot[\kappa_a\, \bm{B}\bb{\bm{x}}  - \zeta_a \bm{E}\bb{\bm{x}} \,],
\end{eqnarray}
which also transforms according to Poincar\'e symmetries described by the extended Poincar\'e group \cite{WuTung}.
In particular, it involves interactions with external vector fields $A_{\mu} = (A_0\bb{\bm{x}},\,\bm{A} \bb{\bm{x}})$, non-minimal couplings with magnetic and electric fields, $\bm{B}\bb{\bm{x}}$ and $\bm{E}\bb{\bm{x}}$, as well as pseudovector field interactions $A_{\mu 5}\sim(q\bb{\bm{x}},\,\bm{W}\bb{\bm{x}})$, and scalar and pseudoscalar field couplings through $\phi_S\bb{\bm{x}}$  and $m\bb{\bm{x}}$, respectively.
Depending on the complexity of the Dirac-like interactions, the computation of quantum information quantifiers can be extremely simplified through the properties identified along this section.

In the next section, as a standard application, our analysis shall be constrained to the phase-space description of a charged fermion trapped by a magnetic field, for which the computation of quantum purity and relative entropy quantifiers shall be now immediate. 

\section{Phase-space description of a charged fermion trapped by a magnetic field}

\hspace{1 em} The Dirac Hamiltonian for a charged fermion of mass $m$ trapped by a magnetic field, ${\bf B} = \mbox{\boldmath$\nabla$} \times {\bf A}$, is given by
\begin{equation}
H_{\rm B}= \mbox{\boldmath$\alpha$} \cdot  ({\bf p} + (-1)^r\, e  {\bf A}) + \beta m,
\label{eq0}
\end{equation}
where $e$ is the positive unit of charge, $r = 1(2)$ for positive(negative) intrinsic parity states, and $\hbar$ and $c$ have been set equal to unity. 
The stationary eigenstates of the above Hamiltonian can be systematically identified by the state vector, $\psi$, as
\begin{equation}
\psi = e^{- i \, E\, t} \left( \begin{array}{c} \phi_a \\ \phi_b \end{array}
\right),
\label{eq1}
\end{equation}
where $E$ describes the energy eigenvalues, and the $\psi$ spinor structure is decoupled into $2$-component spinors, $\phi_{a,b}$.
In the coordinate representation, with ${\bf p} \equiv -i \mbox{\boldmath$\nabla$}$, the corresponding eigenvalue equation can be separated into two coupled equations as
\begin{eqnarray}
(E-m)\phi_a &=& \mbox{\boldmath$\sigma$} \cdot (-i \mbox{\boldmath$\nabla$} +(-1)^r\, e  {\bf A}) \phi_b , 
\label{eq1}\\*
(E+m) \phi_b &=& \mbox{\boldmath$\sigma$} \cdot (-i \mbox{\boldmath$\nabla$} +(-1)^r\, e  {\bf A}) \phi_a ,
\label{eq2}
\end{eqnarray}
with $\mbox{\boldmath$\sigma$} = (\sigma_x,\,\sigma_y,\,\sigma_z)$.

Substituting the expression for $\phi_b$ from Eq.~(\ref{eq2}) into Eq.~(\ref{eq1}), one has
\begin{eqnarray}
(E^2 - m^2)\phi_a &=& \Big[ \mbox{\boldmath$\sigma$}  \cdot (-i\mbox{\boldmath$\nabla$}  +(-1)^r\, e {\bf A}) 
\Big]^2 \phi_a \nonumber\\
&=& \Big[ -\mbox{\boldmath$\nabla$} ^2 + (e {\mathcal B})^2 x^2 -(-1)^r\, e {\mathcal B}(2i\,x
{\partial\over\partial y} - \sigma_z) \Big] \phi_a,
\label{eq3} 
\end{eqnarray}
where, among several gauge-allowed possibilities (not relevant at this point), the vector potential has been identified by ${\bf A} = \mathcal{B}\,x\, \hat{\bf y}$, which gives rise to a uniform magnetic field of magnitude ${\mathcal B}$ along the $z$-direction.
It leads to simultaneous eigenstates of $H_{\rm B}$, $p_x$ and $p_y$ expressed in terms of the $2$-component spinor, $\phi_a$, written 
as
\begin{equation}
\phi_a \equiv \phi^{\pm}_a({\bf x}) = \exp[i(k_y y + k_z z)] \mathcal{F}_{\pm}( x )\, \chi_{\pm}, 
\label{eq4}
\end{equation}
where $\chi_{\pm}$ are the unitary $2$-component spinor eigenvectors of $\sigma_z$, with respective eigenvalues $\pm 1$, and $\mathcal{F}_{\pm}( x )$ satisfy
\begin{eqnarray}
{d^2\mathcal{F}_{\pm} \over dx^2} - (e\,{\mathcal B}\,x + (-1)^r k_y)^2 \mathcal{F}_{\pm} + (E^2 - m^2 -
k_z^2 \mp (-1)^r e {\mathcal B}) 
\mathcal{F}_{\pm} = 0 \,,
\label{eq5}
\end{eqnarray}
obtained from Eq.~(\ref{eq3}), which can be put into the form of the Hermite's equation,
\begin{eqnarray}
\left[ {d^2 \over ds_r^2} -s_r^2 + \zeta_{r,\pm} \right] \mathcal{F}_{\pm}(s_r) = 0 \,,
\label{eq6}
\end{eqnarray}
for
\begin{equation}
s_r = \sqrt{e {\mathcal B}} \left( x + (-1)^r{k_y \over e{\mathcal B}}
\right),
\label{222}
\end{equation}
and
\begin{equation}
\zeta_{r,\pm} = {E^2 - m^2 - k_z^2 \mp (-1)^r e {\mathcal B} \over e{\mathcal B}}.
\label{333}
\end{equation}

Provided $\zeta_{r,\pm}=2n+1$ for $n=0,\,1,\,2,\,\dots$, which results into the energy
eigenvalues 
\begin{equation}
E^2_{n,\pm} = m^2 + k_z^2 + \left[(2n+1) \mp (-1)^r\right] e {\mathcal B},
\label{EE}
\end{equation}
the solutions $\mathcal{F}_{\pm}(s)$ are then identified by 
\begin{equation}
\mathcal{F}_{\pm}(s) = \mathcal{F}_{n}(s) = \left( {\sqrt{e{\mathcal B}} \over n! \, 2^n \sqrt{\pi}} \,
\right)^{1/2} e^{-s^2/2} H_{n}(s),
\end{equation}
where $H_n$ are the Hermite polynomials of order $n$ such that $\mathcal{F}_{n}$ satisfy orthonormalization and completeness relations given respectively by
\begin{equation}
\int_{_{-\infty}}^{^{+\infty}}\hspace{-.5 cm}  \mathcal{F}_n (s) \mathcal{F}_m (s)\,ds = \sqrt{e{\mathcal B}} 
\,\,\,\delta_{nm},
\label{444}
\end{equation}
and
\begin{equation}
\sum_n \mathcal{F}_n(s) \mathcal{F}_n(s') = \sqrt{e{\mathcal B}} \;
\delta(s -s') = \delta (x-x'). 
\label{555}
\end{equation}

The above expression for the energy eigenvalues has an infinite degeneracy with respect to the continuous variable $k_y$ and it also exhibits a $2$-fold degeneracy between the levels with quantum numbers $n-1$ and $+(-)$, and $n$ and $-(+)$,  and for positive(negative) intrinsic parity states ($r = 1(2))$\footnote{It is a consequence of the generalized helicity $\mbox{\boldmath$\sigma$} \cdot  ({\bf p} + (-1)^r\, e  {\bf A})$ which commutes with the Hamiltonian.}, i. e.
\begin{eqnarray}
E^2_{n-1,+} = E^2_{n,-} &=& m^2 + k_z^2 + 2n\, e {\mathcal B},\qquad \mbox{for $r=1$,}\nonumber\\
 E^2_{n-1,-} = E^2_{n,+}  &=& m^2 + k_z^2 + 2n\, e {\mathcal B},\qquad \mbox{for $r=2$.}\nonumber
\label{666}
\end{eqnarray}
Adopting the value of
\begin{eqnarray}
\pm E_{n} &=& \pm \sqrt{m^2 + k_z^2 + 2n\,e{\mathcal B}}
\label{777}
\end{eqnarray}
as the energy of the $n$-th Landau level related to positive and negative intrinsic parity solutions, after some straightforward mathematical manipulations, the complete form of the eigenspinor solutions for the Hamiltonian Eq.~(\ref{eq1}) can be finally expressed by
\begin{equation}
\exp[i((-1)^r E_{n}\,t + k_y y + k_z z)] u^{\pm}_{n,r}(s_r),
\label{1010}
\end{equation}
with 
\begin{eqnarray}
u^+_{n,1}(s_1) = \sqrt{\eta_{n}}\left( \begin{array}{c} 
\mathcal{F}_{n-1}(s_1) \\ 0 \\ 
A_{n}\, \mathcal{F}_{n-1}(s_1) \\
-B_{n}\, \mathcal{F}_{n} (s_1) 
\end{array} \right), \quad 
u^-_{n,1}(s_1) = \sqrt{\eta_{n}}\left( \begin{array}{c} 
0 \\ \mathcal{F}_{n} (s_1) \\
-B_{n}\,
\mathcal{F}_{n-1}(s_1) \\ 
-A_{n}\,\mathcal{F}_{n}(s_1) 
\end{array} \right), \quad 
\end{eqnarray}
and
\begin{eqnarray}
u^-_{n,2}(s_2) = \sqrt{\eta_{n}}\left( \begin{array}{c} 
A_{n}\,
\mathcal{F}_{n-1}(s_2) \\ 
B_{n}\,
\mathcal{F}_{n} (s_2)  \\ 
\mathcal{F}_{n-1}(s_2) \\ 0
\end{array} \right), \qquad 
u^+_{n,2}(s_2) = \sqrt{\eta_{n}}\left( \begin{array}{c} 
B_{n}\,
\mathcal{F}_{n-1}(s_2) \\ 
-A_{n}\,\mathcal{F}_{n}(s_2)  \\ 
0 \\ \mathcal{F}_{n} (s_2)
\end{array} \right),
\label{9999}
\end{eqnarray}
where
\begin{equation}
 A_{n}\, = \frac{k_z}{E_{n} +m},\,\,B_{n}\, = \frac{\sqrt{2n\,e\mathcal{B}}}{E_{n} +m},\,\,\,\mbox{and} \,\,\, \eta_{n}=\frac{E_{n} +m}{2E_{n}}.
\label{nova}\end{equation}

Considering that the multiplicative exponential factor from Eq.~(\ref{1010}) does not affect the computation of pure state Wigner functions\footnote{For gauge field interactions, as it is supposed from electrons trapped by magnetic fields, slowly varying external ($c$-number) fields can be expanded as $F^{\mu\nu}({x}+{x}') \approx F^{\mu\nu}({x})+ x_{\lambda}\partial^{\lambda}F^{\mu\nu}({x})$ where the $c$-number field is homogeneous if the derivative term can be neglected. In the case where the gauge field is summarized by the variable $\mathcal{B}$ ($z$-direction), in a typically homogeneous configuration, the associated Wigner function coincides with the naive gauge dependent definition (cf. Sec. 4.1 of Ref. \cite{1986}) the momentum in the exponential of the Fourier transform is replaced by $({\bf p} + (-1)^r\, e  {\bf A})$ which, for the field configurations considered here, leads to the simple redefinition of the quantum mechanical momentum component $k_y$ as $\pi_y = k_y + (-1)^r\, e\, \mathcal{B} x)$, which is averaged out by $(y,\,\pi_y)$ integrations. Gauge invariance is not relevant for the quantum information $1$-dim analysis here performed.}, the following analysis can be reduced to an effective $1$-dim calculation where the relevant quantities are obtained from the spinors $u^{\pm}_{n,r}(s_r)$. Once the spinor probability distributions for the pure eigenstates have been put in the matrix form, $u^{\dagger\pm}_{n,r}(s_r)\,\gamma_{0}\,\,u^{\pm}_{n,r}(s_r)$, the following set of pure state Dirac-like Wigner functions can be obtained (cf. the $1$-dim reduced version of Eq.~(\ref{122223})),
\begin{eqnarray}
\omega^+_{n,1}(s_1,\,k_x) &=&\eta_{n}\left( \begin{array}{cccc} 
\mathcal{L}^{(1)}_{n-1} & 0 & -A_{n}\,\mathcal{L}^{(1)}_{n-1} & B_{n}\,\mathcal{M}^{(1)}_{n}  \\ 
0 & 0& 0& 0\\
A_{n}\,\mathcal{L}^{(1)}_{n-1} & 0 & -A^2_{n}\,\mathcal{L}^{(1)}_{n-1} & A_{n}B_{n}\,\mathcal{M}^{(1)}_{n}  \\ 
-B_{n}\,\mathcal{M}^{(1)}_{n} & 0 & A_{n}B_{n}\,\mathcal{M}^{(1)}_{n}& -B^2_{n}\,\mathcal{L}^{(1)}_{n}    
\end{array} \right), 
\end{eqnarray}
\begin{eqnarray}
\omega^-_{n,1}(s_1,\,k_x) &=& \eta_{n}\left( \begin{array}{cccc} 
0 & 0& 0& 0\\
0 & \mathcal{L}^{(1)}_{n}  & B_{n}\,\mathcal{M}^{(1)}_{n}&  A_{n}\,\mathcal{L}^{(1)}_{n}  \\ 
0 & -B_{n}\,\mathcal{M}^{(1)}_{n} & -B^2_{n}\,\mathcal{L}^{(1)}_{n-1}& -A_{n}B_{n}\,\mathcal{M}^{(1)}_{n}  \\  
0 & -A_{n}\,\mathcal{L}^{(1)}_{n} & -A_{n}B_{n}\,\mathcal{M}^{(1)}_{n}& -A^2_{n}\,\mathcal{L}^{(1)}_{n}   \\ 
\end{array} \right),
\end{eqnarray}
\begin{eqnarray}
\omega^-_{n,2}(s_2,\,k_x) &=& \eta_{n}\left( \begin{array}{cccc} 
A^2_{n} \mathcal{L}^{(2)}_{n-1} & A_{n}B_{n}\,\mathcal{M}^{(2)}_{n}  & -A_{n}\,\mathcal{L}^{(2)}_{n-1} & 0  \\ 
A_{n}B_{n}\,\mathcal{M}^{(2)}_{n}  & B^2_{n}\,\mathcal{L}^{(2)}_{n}  & -B_{n}\,\mathcal{M}^{(2)}_{n} & 0  \\ 
A_{n}\,\mathcal{L}^{(2)}_{n-1} & B_{n}\,\mathcal{M}^{(2)}_{n} & -\mathcal{L}^{(2)}_{n-1}  & 0  \\ 
0 & 0& 0& 0
\end{array} \right), 
\end{eqnarray}
\begin{eqnarray}
\omega^+_{n,2}(s_2,\,k_x) &=& \eta_{n}\left( \begin{array}{cccc} 
B^2_{n} \mathcal{L}^{(2)}_{n-1} & - A_{n}B_{n}\,\mathcal{M}^{(2)}_{n}  & 0 & -B_{n}\,\mathcal{M}^{(2)}_{n} \\ 
- A_{n}B_{n}\,\mathcal{M}^{(2)}_{n}  & A^2_{n}\,\mathcal{L}^{(2)}_{n}  & 0 &A_{n}\,\mathcal{L}^{(2)}_{n}\\ 
0 & 0& 0& 0\\
B_{n}\,\mathcal{M}^{(2)}_{n} & -A_{n}\,\mathcal{L}^{(2)}_{n} &0 & -\mathcal{L}^{(2)}_{n} \\ 
\end{array} \right),
\label{9999}
\end{eqnarray}
where
\begin{equation}\label{norr01}
\mathcal{L}^{(r)}_{n} \equiv \mathcal{L}_{n}(s_r,\,k_x) = (-1)^{n} \sqrt{e\mathcal{B}}\,\exp\left[-(s_r^2 + k^2_x)\right]\, {L}_{n}\left[2(s_r^2 + k^2_x)\right],
\end{equation}
with $L_{n}$ corresponding to the Laguerre polynomials of order $n$, such that
\begin{equation}\label{norr0w}
\int_{_{-\infty}}^{^{+\infty}}\hspace{-.5 cm}dx\,\int_{_{-\infty}}^{^{+\infty}}\hspace{-.5 cm}dk_x \,\mathcal{L}_{n}(s_r,\,k_x) = 1,\,\, (\mbox{with}\quad ds_r = \sqrt{e\mathcal{B}}\, dx),
\end{equation}
and
\begin{eqnarray}
\mathcal{M}^{(r)}_{n} &\equiv& \mathcal{M}_{n}(s_r,\,k_x)\nonumber\\&=& \frac{\sqrt{n \, e\mathcal{B}}}{2^{n-\frac{3}{2}}n!\sqrt{\pi}}
\exp\left[-s_r^2\right]\int_{_{-\infty}}^{^{+\infty}}\hspace{-.5 cm}du\,\exp\left[2\,i\, k_x\, u\right]\exp\left[-u^2\right]\times\nonumber\\
&&\qquad\qquad\qquad\qquad\qquad\frac{1}{2}\big{\{}H_{n}(s_r+u)\,H_{n-1}(s_r-u)+H_{n-1}(s_r+u)H_{n}(s_r-u) \big{\}}\nonumber\\
&=& \frac{\sqrt{ e\mathcal{B}}}{2^{n+\frac{1}{2}}n!\sqrt{n\pi}}
\exp\left[-s_r^2\right]\int_{_{-\infty}}^{^{+\infty}}\hspace{-.5 cm}du\,\exp\left[2\,i\, k_x\, u\right]\exp\left[-u^2\right]
\frac{d}{ds_r}\big{\{}H_{n}(s_r+u)H_{n}(s_r-u) \big{\}}\nonumber\\
&=& (-1)^{n}\sqrt{\frac{ e\mathcal{B}}{2^3 n}}\,\exp\left[-(s_r^2 + k^2_x)\right] \left(\frac{d}{ds_r}{L}_{n}\left[2(s_r^2 + k^2_x)\right]\right),
\end{eqnarray}
which is an odd function of $s_r$ and therefore
\begin{equation}
\int_{_{-\infty}}^{^{+\infty}}\hspace{-.5 cm}dx\,\int_{_{-\infty}}^{^{+\infty}}\hspace{-.5 cm}dk_x \,\mathcal{M}_{n}(s_r,\,k_x) = \int_{_{-\infty}}^{^{+\infty}}\hspace{-.5 cm}ds_r\int_{_{-\infty}}^{^{+\infty}}\hspace{-.5 cm}dk_x\, \mathcal{M}_{n}(s_r,\,k_x) = 0.
\end{equation}
Besides the above orthonormalization integrations, one should notice that \begin{eqnarray}
\rho^{^{+(-)}}(s_{1(2)},\,k_x) &=& Tr\left[\omega^{+(-)}_{n,1(2)}\,\gamma_{0}\,\right] = 
\eta_n \left[(1+A_{n}^2)\mathcal{L}^{(1(2))}_{n-1} + B^2_{n} \mathcal{L}^{(1(2))}_{n}\right],\nonumber\\
\rho^{^{-(+)}}(s_{1(2)},\,k_x) &=& Tr\left[\omega^{-(+)}_{n,1(2)}\,\gamma_{0}\,\right] = 
\eta_n \left[(1+A_{n}^2)\mathcal{L}^{(1(2))}_{n} + B^2_{n} \mathcal{L}^{(1(2))}_{n-1}\right],
\label{1111}
\end{eqnarray}
which, as expected, once $s_r$ is replaced by its $x$ dependent expression,
result into
\begin{equation}
\int_{_{-\infty}}^{^{+\infty}}\hspace{-.5 cm}dx\,\int_{_{-\infty}}^{^{+\infty}}\hspace{-.5 cm}dk_x 
\rho^{^{\pm}}(s_{r}(x),\,k_x) = \eta_n\left[(1+A_{n}^2) + B^2_{n}\right] = 1.
\end{equation}

From the above quantities, quantum purity and relative linear entropies can be straightforwardly computed.
Of course, the quantum purity for pure eigenstates results into unity\footnote{In this case, one has
\begin{eqnarray}
\mathcal{P} &=& \frac{2\pi}{\sqrt{e\mathcal{B}}}\int_{_{-\infty}}^{^{+\infty}}\hspace{-.5 cm}dx\,\int_{_{-\infty}}^{^{+\infty}}\hspace{-.5 cm}dk_x 
Tr\left[\omega^{+}_{n,1(2)}\,\gamma_{0}\,\omega^{+}_{n,1(2)}\,\gamma_{0}\,\right]
=
\frac{2\pi}{\sqrt{e\mathcal{B}}}\int_{_{-\infty}}^{^{+\infty}}\hspace{-.5 cm}dx\,\int_{_{-\infty}}^{^{+\infty}}\hspace{-.5 cm}dk_x 
Tr\left[\omega^{-}_{n,2(1)}\,\gamma_{0}\,\omega^{-}_{n,2(1)}\,\gamma_{0}\,\right]
\nonumber\\ &=&
\frac{2\pi}{\sqrt{e\mathcal{B}}}\eta_{n}^2
\int_{_{-\infty}}^{^{+\infty}}\hspace{-.5 cm}dx\,\int_{_{-\infty}}^{^{+\infty}}\hspace{-.5 cm}dk_x
\left[\left(1+A_{n}^2\right)^2\mathcal{L}^2_{n-1(n)}  + 2B^2_{n}\left(1+A_{n}^2\right)\mathcal{M}^2_{n}  +B^4_{n}\mathcal{L}^2_{n(n-1)} \right] =
\nonumber\\ &=&
\eta_{n}^2\left[(1+A_{n}^2) + B^2_{n}\right]^2 = 1,
\end{eqnarray}
where the explicit dependence on $(s_r,\,k_x)$ has been suppressed from the notation, and the multiplicative parameter $1/\sqrt{e\mathcal{B}}$ appears due to the choice of the normalization conditions.}, $\mathcal{P} =1$.
The relative linear entropies related to discrete spin-parity ($SP$) and continuous ($1$-dim reduced) position-momentum $(x,\,k_x)$ degrees of freedom are respectively given by
\begin{eqnarray}
\mathcal{I}^{SP} &=& 1 - Tr\left[\left(\gamma_{0}\,\int_{_{-\infty}}^{^{+\infty}}\hspace{-.5 cm}dx\,\int_{_{-\infty}}^{^{+\infty}}\hspace{-.5 cm}dk_x \,\omega^{\pm}_{n,r}(s_r,\,k_x)\right)^2\right]=
1 - \eta_{n}^2\left[(1+A_{n}^2)^2 + B^4_{n}\right],
\end{eqnarray}
and
\begin{eqnarray}
\mathcal{I}_{\{x,k_x\}} &=& 1 - \frac{2\pi}{\sqrt{e\mathcal{B}}}\int_{_{-\infty}}^{^{+\infty}}\hspace{-.5 cm}dx\,\int_{_{-\infty}}^{^{+\infty}}\hspace{-.5 cm}dk_x 
\left(\rho^{^{\pm}}(s_r(x),\,k_x)\right)^2=
1 - \eta_{n}^2\left[(1+A_{n}^2)^2 + B^4_{n}\right],
\end{eqnarray}
which exhibit coincident values.

\begin{figure}
\vspace{-1.5cm}\includegraphics[scale=0.53]{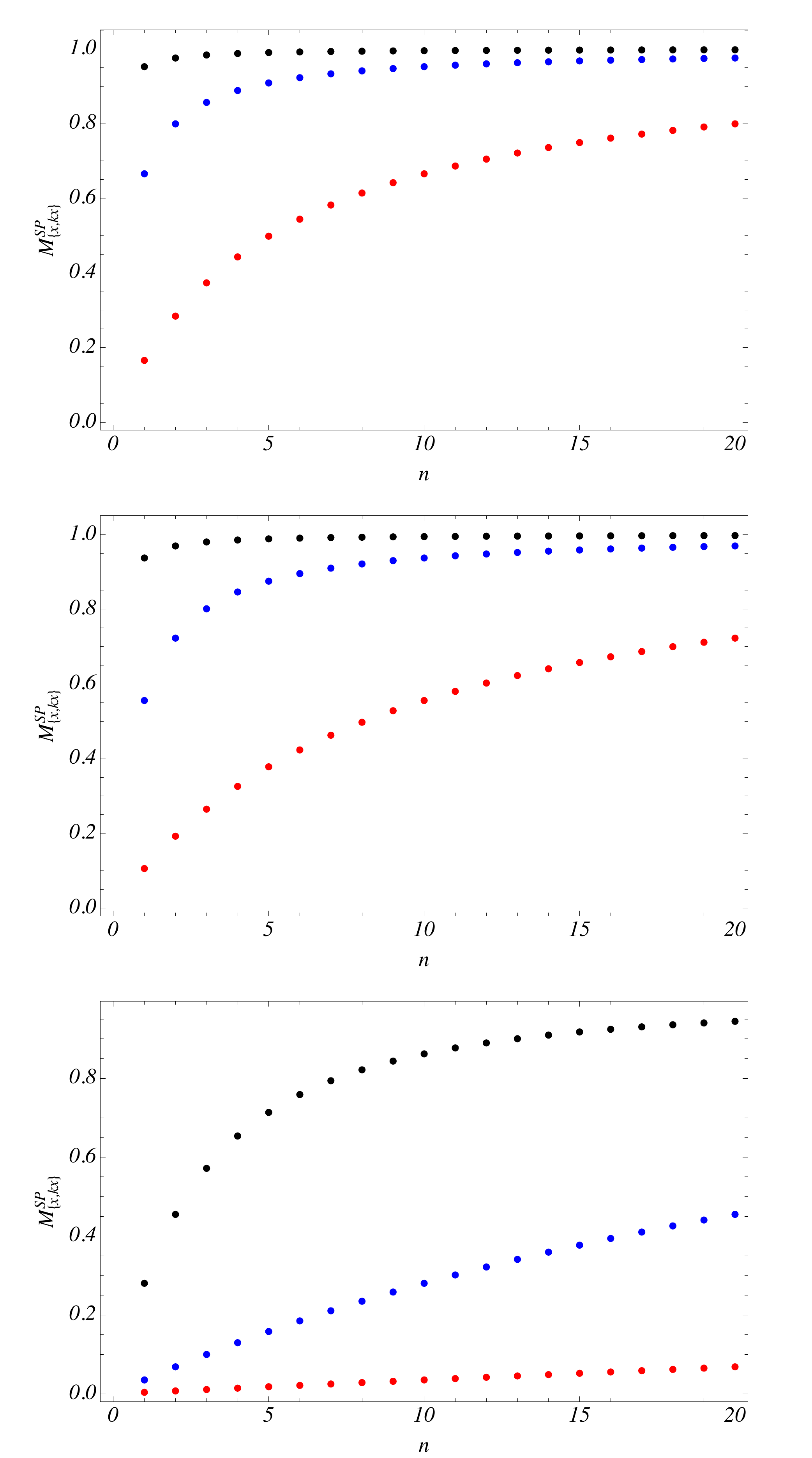}
\renewcommand{\baselinestretch}{.85}
\caption{\footnotesize{
(Color online) Mutual information $M^{SP}_{\{x,k_x\}}$ as function of $n$, for non-relativistic ($\kappa = 0.01$ (first plot)) relativistic ($\kappa = 1$ (second plot)) and and ultra-relativistic ($\kappa = 100$ (third plot)) regimes.
The plots are for $e\mathcal{B}/m^2 = 0.1$ (red dots), $1$ (blue dots) and $10$ (black dots).}}
\label{f1}
\end{figure}

Finally, the complete expression for the mutual information between spin-parity ($SP$) and position-momentum $(x,\,k_x)$ degrees of freedom, in terms of the quantum number $n$ and the interacting parameter $e\mathcal{B}$ (cf. Eq.~(\ref{nova})) is explicitly given by
\begin{eqnarray}
M^{SP}_{\{x,k_x\}}(n) &=& \mathcal{I}_{\{x,k_x\}} + \mathcal{I}^{SP} - (1 - \mathcal{P})\nonumber\\
&=& 
4\eta_{n}^2(1+A_{n}^2)\, B^2_{n}\nonumber\\
&=& \frac{2n\,e\mathcal{B}}{m^2 + k_z^2 + 2n\, e {\mathcal B}}\left[1+ \frac{k_z^2}{(\sqrt{m^2 + k_z^2 + 2n\, e {\mathcal B}}+m)^2}\right]\nonumber\\
&=& \frac{2n\,\epsilon}{1 + \kappa + 2n\, \epsilon }\left[1+ \frac{\kappa}{(\sqrt{1 + \kappa + 2n\, \epsilon}+1)^2}\right],
\end{eqnarray}
with $\epsilon = e\mathcal{B}/m^2$ and $\kappa = k_z^2/m^2$, and which is depicted in Fig.~\ref{f1}.
Assuming that the kinematical regime is driven by $\kappa$ (in the case where $k_y$ is not relevant), one notices that $M^{SP}_{\{x,k_x\}}$ is highly suppressed for ultra-relativistic regimes and it vanishes for $\kappa \to \infty$. For non-relativistic regimes ($\kappa \to 0$), one has $M^{SP}_{\{x,k_x\}}(n) \to {2n\,\epsilon}/(1  + 2n\, \epsilon)$, which saturates to unity for large $n$ values.

\begin{figure}
\vspace{-1.5cm}\includegraphics[scale=0.83]{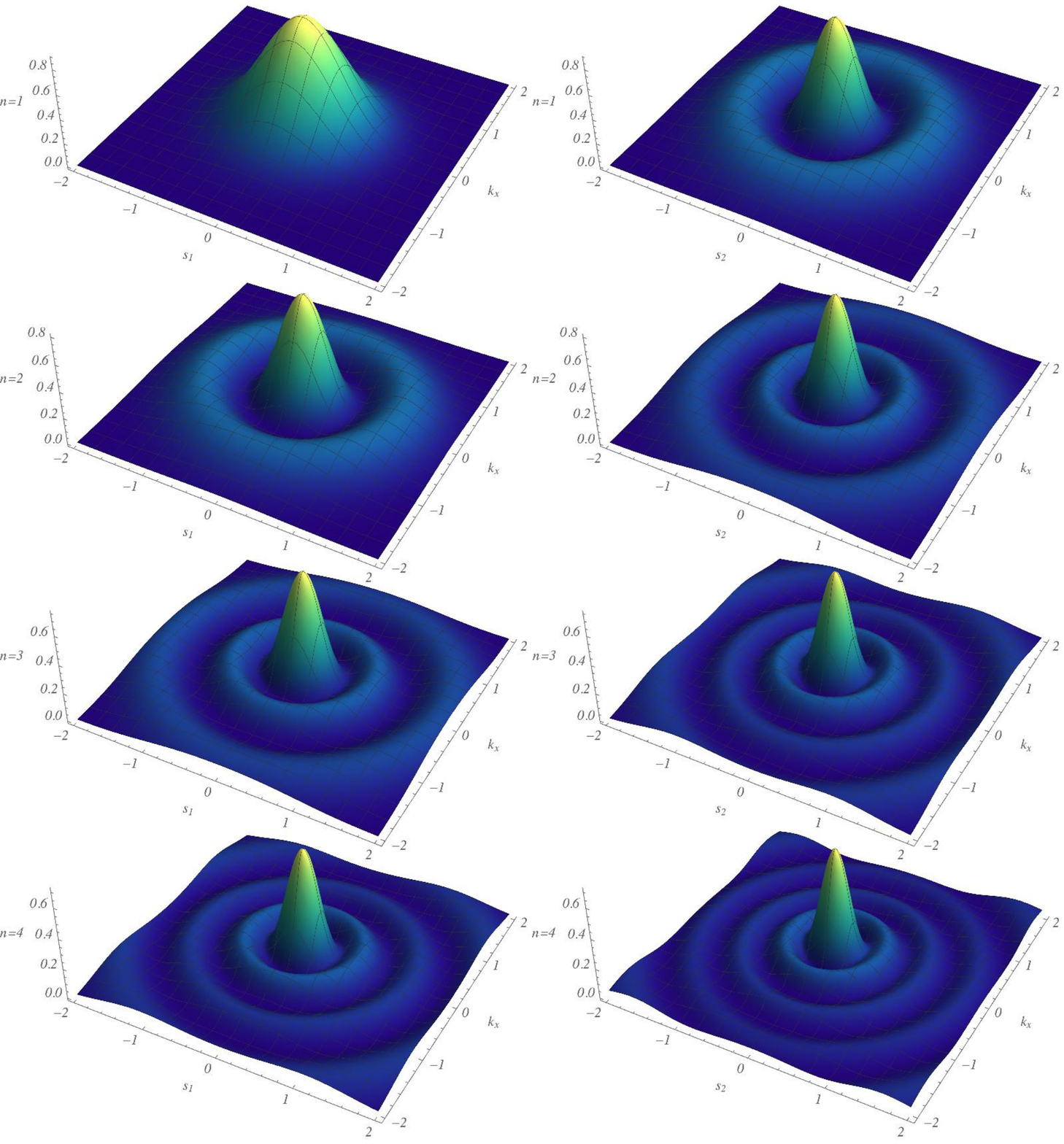}
\renewcommand{\baselinestretch}{.85}
\vspace{-2.5cm}\caption{\footnotesize{
(Color online) Phase-space pattern of dimensionless (cf. normalization condition from Eqs.~(\ref{norr01})-(\ref{norr0w})) quantum purity for simultaneous spin-up(down) positive(negative) parity states, $(1/{e\mathcal{B}}) \mathcal{P}^{+(-)}(s_{1(2)},\,k_x)$ (first row), and spin-up(down) negative(positive) intrinsic parity states, $(1/{e\mathcal{B}}) \mathcal{P}^{-(+)}(s_{1(2)},\,k_x)$ (second row).
The two sets of plots are for quantum numbers $n$ from $1$ to $4$, with $\kappa = k_z^2/m^2 = 1$ and $\epsilon = e\mathcal{B}/m^2 =1$.}}
\label{f2}
\end{figure}

To summarize, $M^{SP}_{\{x,k_x\}}(n)$ measures how much information is communicated, on average, in spin-parity Hilbert space about the phase-space, and vice-versa, evidently, due to the coupling property of the magnetic field. Free fermions do not exhibit such a mutual influence.
Of course, the way it has been obtained accounts for all the subtleties of the Dirac equation solutions in the phase-space. 
One of these subtle aspects can also be identified in terms of the local phase-space behavior of the (non integrated) quantum purity 
\begin{eqnarray}
\mathcal{P}^{+(-)}(s_{1(2)},\,k_x) &=& 
Tr\left[\omega^{+(-)}_{n,1(2)}\,\gamma_{0}\,\omega^{+(-)}_{n,1(2)}\,\gamma_{0}\,\right]
\nonumber\\&=&
\eta_{n}^2
\left[\left(1+A_{n}^2\right)^2\mathcal{L}^2_{n-1}(s_{1(2)},\,k_x) + \right. \nonumber\\ &&\qquad\qquad\qquad \left. 2B^2_{n}\left(1+A_{n}^2\right)\mathcal{M}^2_{n}(s_{1(2)},\,k_x)  +B^4_{n}\mathcal{L}^2_{n}(s_{1(2)},\,k_x) \right],\\
\mathcal{P}^{-(+)}(s_{1(2)},\,k_x) &=& 
Tr\left[\omega^{-(+)}_{n,1(2)}\,\gamma_{0}\,\omega^{-(+)}_{n,1(2)}\,\gamma_{0}\,\right]
\nonumber\\ &=&
\eta_{n}^2
\left[\left(1+A_{n}^2\right)^2\mathcal{L}^2_{n}(s_{1(2)},\,k_x)  + \right. \nonumber\\ &&\qquad\qquad\qquad \left. 2B^2_{n}\left(1+A_{n}^2\right)\mathcal{M}^2_{n}(s_{1(2)},\,k_x)  +B^4_{n}\mathcal{L}^2_{n-1}(s_{1(2)},\,k_x) \right], 
\end{eqnarray}
which emphasize the discrimination between $\omega^{+(-)}_{n,1(2)}$ and $\omega^{+(-)}_{n,2(1)}$ associated spinor states, as depicted through the phase-space behavior in Fig.~\ref{f2}. In spite of exhibiting the same relative linear entropies, as well as the same mutual information between spin-parity and phase-space coordinates, the plots from Fig.~\ref{f2} certify that such a correspondence is due to an averaged behavior: locally, the purity related to spin-parity components depends on the combination between spin ($\pm$) and parity ($r=1,\,2$) quantum configurations.

\subsection{Spin-parity intrinsic quantum concurrence for charged fermions in a uniform magnetic field}

Considering that several theoretical tools \cite{n022,n023} can be used for the measuring the quantum entanglement, one can preliminarily suppose that, at least for the Dirac spinor structure discussed in the previous section, the entanglement of formation (EoF)\cite{n024} -- the convex-roof extension of the von Neumann entropy of the quantum state $\varrho$, $E_{vN}[\varrho]$ -- introduced as the mean value of the pure-state entanglement, which is minimized over the total number of decompositions of the mixed state $\varrho$ on pure states, $\varrho_k$,
\begin{equation}
E_{EoF} [\varrho] = \mbox{min}_{\varrho_k} \displaystyle \sum_k q_k E_{vN}[\varrho_k],
\end{equation}
can be assumed as the driver measure of quantum entanglement.

For two-qubit states, the EoF is obtained in terms of the quantum concurrence \cite{n024}, $\mathcal{C}[\varrho]$, as
\begin{eqnarray}
E_{EoF} [\varrho] &=& \mathcal{E}\left[ \frac{1 - \sqrt{1 - \mathcal{C}^2[\varrho]}}{2}\right],
\end{eqnarray}
with $\mathcal{E}[\lambda] =- \lambda \log_2 \lambda - (1-\lambda)\log_2 (1-\lambda)$, and where $\mathcal{C}[\varrho]$ can straightforwardly identified as an entanglement quantifier given by defined as \cite{n024}
\begin{equation}
\mathcal{C}[\varrho] = \mbox{max}\{ \omega_1 - \omega_2 - \omega_3 - \omega_4 \, , \,0 \},
\end{equation}
where $\omega_1 > \omega_2 > \omega_3 > \omega_4$ are the eigenvalues of the operator $\sqrt{\, \sqrt{\varrho} \, (\sigma_y \otimes \sigma_y) \varrho^\ast (\sigma_y \otimes \sigma_y) \, \sqrt{\varrho}\, }$. 

In the scope of the spin-parity intrinsic entanglement for Dirac spinor solutions, for a generic pure state described by $\vert w \rangle\langle w\vert$, the qubit-flip operation returns  $\vert \widetilde{w} \rangle\langle \widetilde{w}\vert$, with
\begin{equation}
\vert \widetilde{w} \rangle = \sigma^{(1)}_y\otimes\sigma^{(2)}_y\vert {w}^* \rangle,
\end{equation}
where ``$*$'' denotes the complex conjugation operator.
In this case, by identifying the associated density matrix with $\varrho = \vert w \rangle\langle w\vert$, one has the quantum concurrence given by
\begin{equation}
\mathcal{C}[\varrho] = \sqrt{\langle w  \vert \widetilde{\rho}\vert w\rangle} = \vert \langle w \vert \widetilde{w} \rangle\vert = \sqrt{Tr [ \varrho \widetilde{\varrho}]},
\end{equation}
with $\widetilde{\varrho} = \vert \widetilde{w} \rangle \langle \widetilde{w} \vert$.
Once that $\varrho$ can be generically identified by
\begin{equation}
\varrho = \frac{1}{4} \left[ I + (\bm{\sigma}^{(1)} \otimes I^{(2)}) \cdot \bm{a} + (I^{(1)} \otimes \bm{\sigma}^{(2)}) \cdot \bm{b}  + \displaystyle \sum_{i,j = 1}^3 t_{ij} (\sigma_i^{(1)} \otimes \sigma_j^{(2)}) \right],
\end{equation}
where $t_{ij}$ are the elements of the correlation matrix, $\bm{T}$, and $\bm{a}$ and $\bm{b}$ are the Bloch vectors of the corresponding subsystems, for pure states, where $a^2 = b^2$, and the concurrence is simply given by
\begin{equation}
\mathcal{C}[\varrho] = \sqrt{1 - a^2}.
\end{equation}

\begin{figure}
\vspace{-1.5cm}\includegraphics[scale=0.83]{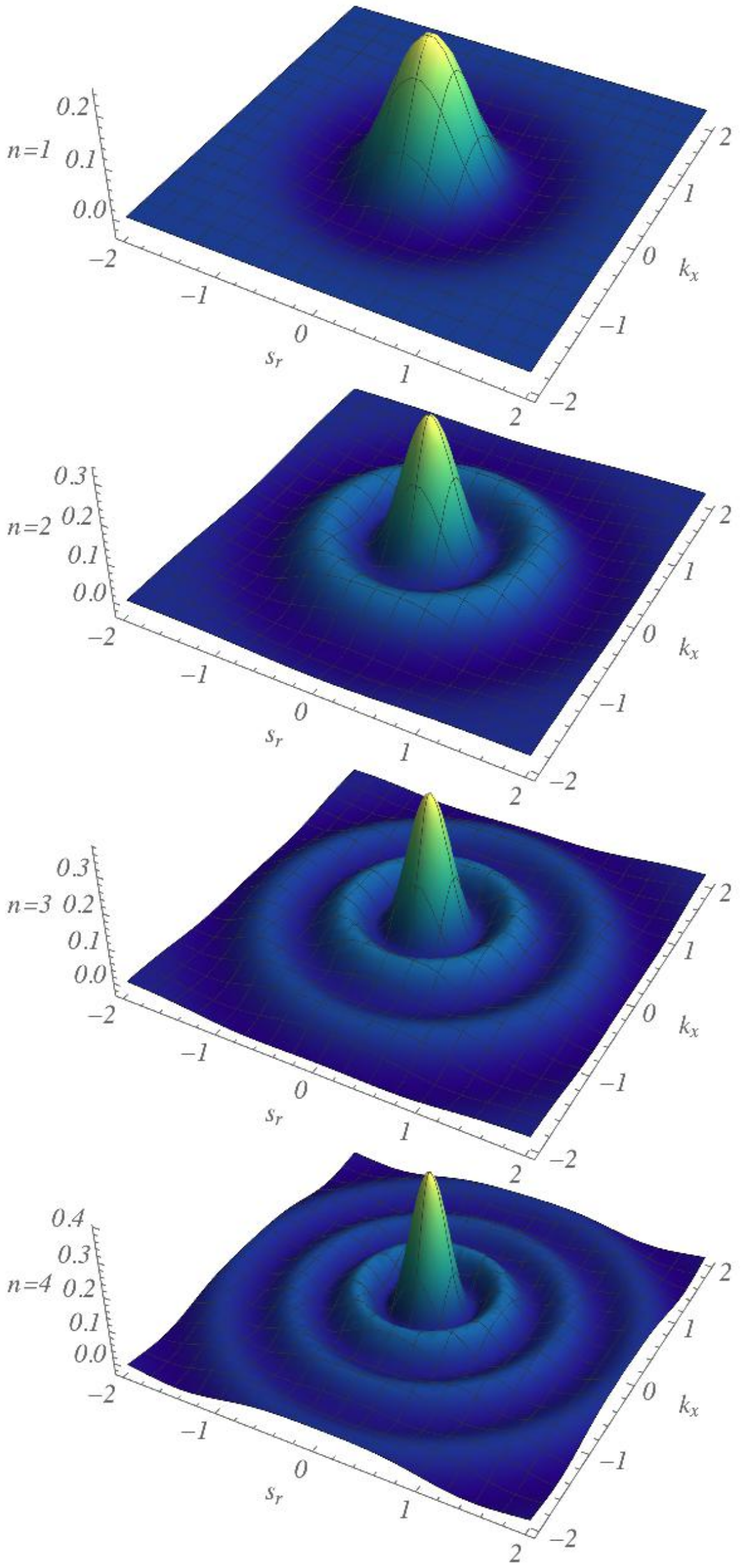}
\renewcommand{\baselinestretch}{.85}
\vspace{-2.5cm}\caption{\footnotesize{
(Color online) Phase-space pattern of the spin-parity dimensionless (dimensionless cf. normalization condition from Eqs.~(\ref{norr01})-(\ref{norr0w})) quantum concurrence $(1/{e\mathcal{B}})\mathcal{C}^2[\omega^{\pm}_{n,r}]$.
The set of plots are for quantum numbers $n$ from $1$ to $4$, with $\kappa = k_z^2/m^2 = 1$ and $\epsilon = e\mathcal{B}/m^2 =1$.}}
\label{f3}
\end{figure}

Once one has the phase-space structure for Dirac spinors as identified through the Wigner functions from the previous section, the intrinsic quantum correlation between spin and parity can be straightforwardly evaluated in terms of the quantum concurrence of $\gamma_{0}\,\omega^{\pm}_{n,r}$ correspondent to $\varrho$, given a natural correspondence with $\widetilde{\varrho}$ identified by $\widetilde{\gamma^{0}\,\omega^{\pm}_{n,r}} = (-i\gamma^{2})\gamma^{0}\,\omega^{\pm}_{n,r}\,(-i\gamma^{2}) $.
To be more specific, given that the qubit-flip operator $\sigma^{(P)}_y\otimes\sigma^{(S)}_y$ is identified by $-i\,\gamma^{\2}$, the corresponding qubit-flipped operators are identified by 
$\widetilde{\gamma^{0}\,\omega^{\pm}_{n,r}} = (-1)\gamma^{2}\gamma^{0}\,\omega^{\pm}_{n,r}\,\gamma^{2}$, such that $Tr[\widetilde{\gamma^{0}\,\omega^{\pm}_{n,r}} ] = Tr[{\gamma^{0}\,\omega}^{\pm}_{n,r} ]$, where $(\gamma^{\2})^2 = -1$.
For the calculation of the square of the quantum concurrence, one typically has
\begin{equation}
\mathcal{C}^2[\omega^{\pm}_{n,r}] = Tr[{\gamma^{0}\,\omega}^{\pm}_{n,r}\,\widetilde{\gamma^{0}\,\omega^{\pm}_{n,r}}] =
(-1) Tr[\omega^{\pm}_{n,r}\,\gamma^2\gamma^{0}\,\omega^{\pm}_{n,r}\,\gamma^2\gamma^{0}],
\end{equation}
which, for the magnetically trapped electron(positron) pure state Dirac-Wigner solutions from the previous section, results into
\begin{equation}
\mathcal{C}^2[\omega^{\pm}_{n,r}] = -2\eta_n^2 B_n^2 \mathcal{L}^{(r)}_n\,  \mathcal{L}^{(r)}_{n-1} \equiv -\frac{n\,e\mathcal{B}}{m^2+k_z^2+ 2 n\,e\mathcal{B}} \mathcal{L}^{(r)}_n(s_{r},\,k_x)\,  \mathcal{L}^{(r)}_{n-1}(s_{r},\,k_x)
\end{equation}
which, despite averaging out to zero, exhibits a subtle phase-space localized pattern similar to that for quantum purity, as depicted in Fig.~\ref{f3}.

In both cases, for quantum purity and quantum concurrence, besides the evinced cylindrical symmetry, the number of radial nodes increase with the quantum numbers.
The local characteristic associated to spin-parity degrees of freedom is distributed over all the phase-space, being more concentrated around $k_x =0$ and $x = \pm k_y/e\mathcal{B} \,(s_r = 0)$.

To summarize, a short note on the Gordon decomposition of Dirac density currents is necessary, given its relevance in the interpretation of the Dirac equation as well as in establishing the correspondence between Maxwell-Dirac and Maxwell-Lorentz theories \cite{Caban,Peres,Caban2,Czachor}. 
Through the notation adopted along this work, Dirac density currents are identified by
\begin{eqnarray}
Q^{-1}j^{\mu}(\mathbf{x}) = \overline{\psi}(\mathbf{x})\,\gamma^{\mu}\,{\psi}(\mathbf{x}) &=& \int d^{3}\mathbf{k}\,\,Tr\left[\,\gamma_{\mu}\,\omega(\mathbf{x},\,\mathbf{k};\,t)\right] = 4\,\int d^{3}\mathbf{k}\, \mathcal{V}^{\mu}(\mathbf{x},\,\mathbf{k};\,t),\\
j^{\mu}_{5}(\mathbf{x}) = \overline{\psi}(\mathbf{x})\,\gamma_{5}\,\gamma^{\mu}\,{\psi}(\mathbf{x}) &=& \int d^{3}\mathbf{k}\,\,Tr\left[\,\gamma_{5}\,\gamma_{\mu}\,\omega(\mathbf{x},\,\mathbf{k};\,t)\right] = 4\,\int d^{3}\mathbf{k}\, \mathcal{A}^{\mu}(\mathbf{x},\,\mathbf{k};\,t),
\end{eqnarray}
respectively for the charge (vector) and chiral (axial) current densities, which respectively couple to $A_{\mu} = (A_0\bb{\bm{x}},\,\bm{A} \bb{\bm{x}})$ and $A_{\mu 5}\sim(q\bb{\bm{x}},\,\bm{W}\bb{\bm{x}})$ (cf. Eq.~(\ref{E04T})), and 
\begin{equation}
j^{\mu\nu}(\mathbf{x}) = \overline{\psi}(\mathbf{x})\,\sigma^{\mu\nu}\,{\psi}(\mathbf{x}) = \int d^{3}\mathbf{k}\,Tr\left[\,\sigma^{\mu\nu}\,\omega(\mathbf{x},\,\mathbf{k};\,t)\right] = 4\,\int d^{3}\mathbf{k}\, \mathcal{T}^{\mu\nu}(\mathbf{x},\,\mathbf{k};\,t),
\end{equation}
for the (spin-associated) tensor current densities, from which
\begin{equation}
 \int d^{3}\mathbf{x}\,j^{0k}(\mathbf{x}) =  \int d^{3}\mathbf{x}\,{\psi}^{\dagger}(\mathbf{x})\,\Sigma^k{\psi}(\mathbf{x}) =  \int d^{3}\mathbf{x}\int d^{3}\mathbf{k}\,\,Tr\left[\,\gamma^{\0}\Sigma^{k}\,\omega(\mathbf{x},\,\mathbf{k};\,t)\right] = \langle \Sigma^k \rangle,
\end{equation}
is the spin component averaged contribution to the spin current \cite{Caban,Cabrera}.
As it can be noticed, the averaged values of the above quantities lead to the same variables that contribute to the computation of $\mathcal{P}$ and $\mathcal{C}$, however, neither a general rule can be formulated, nor a specific coincidence for the electrons driven by the Hamiltonian (\ref{eq0}) can be identified even considering that they depend on the same parameters introduced by Eq.~(\ref{nova}).

Finally, despite not being pertinent to the localized static construction developed here, the connection of the above quantities with quantum observables, namely the current associated conductivities, is a little bit more enhanced. 
For charge and spin conductivities, they are achieved through the Kubo linear response theory \cite{Kubo,Cabrera}, in terms of the functional integrations like, 
\begin{equation}
j^{\mu}(\mathbf{x}) = \int_{-\infty}^{+\infty} dt\, K^{\mu\nu}(t - t') \,A_{\nu}(t').
\end{equation}
for instance, for charge currents. In this case, for externally applied fields which are homogeneous in space, the kernel $K^{\mu\nu}(t - t') $ is obtained from the averaged current-current correlation function, as
\begin{equation}
K^{\mu\nu}(t - t') =i\,\hbar^{-1}\,\theta(t - t') \int d^{3}\mathbf{x}'\, 
\langle [j^{\mu}(\mathbf{x}),\,j^{\nu}(\mathbf{x}') ]\rangle_T,
\end{equation}
where $\theta$ is the Heaviside function, and $\langle ... \rangle_T$ corresponds thermal average and which, as mentioned above, is not concerned with intrinsic spin-parity properties fot the stationary pure state solutions here investigated.
In particular, the engendering of protocols associated to some non-stationary behavior, where phenomenological (measurable) currents could be associated to purity and entanglement properties, demands for, at least, two quantum state superposition, which makes the identification of spin-parity contributions much more obscure. In fact, plethora of issues like that are much more closer to magneto-eletronic \cite{Prinz} and spintronic \cite{todos} topics developed in the last decade.

At our stage, one could only superpose different chiral eigenstates, an look for chiral oscillation implications on entanglement, or superpose different parity eigenstates, and look for some connections with the {\em zitterbewegung}. But it is much more influenced by the superposition coefficients than by intrinsic spinor properties, as it has been emphasized in this work, which is constrained to intrinsic spin-parity properties as the ground for next investigations.

\section{Conclusions}

The interpretation of quantum information correlations under the light of relativistic quantum mechanics (or even of quantum field theory) has namely attained to the issues concerned with how spin-spin and spin-momentum entanglement does change under Lorentz boosts, in particular, in the context of describing communication schemes in the relativistic framework \cite{relat01, relat02, relat03, relat04, relat05, relat06, relat07, relat08, relatvedral}.
In order to provide simplified forms for elementary information quantifier tasks relevant to such scenarios, the phase-space information content related to spin-parity and position-momentum degrees of freedom 
were investigated in the context of the (covariant) Wigner formalism.
Analytical expressions for computing quantum purity, relative linear entropies, mutual information between such discrete and continuous degrees of freedom were obtained. Extensions for the computation of spin-parity intrinsic entanglement were also provided.
In particular, the quantifying theoretical tools were applied for describing such an elementary information profile of a charged fermion trapped by a uniform magnetic field. In this case, the phase-space structure was completely obtained in terms of Laguerre polynomials associated to the quantized energy Landau levels.
Assuming that some mathematical manipulability of the Weyl transformed associated quantum states has been identified, our results are shown to be consistent with the grounds for treating Dirac-like Wigner described phase-space quantum systems.

Conclusively, the first step in the systematic computation of the elementary information content of Dirac-like systems exhibiting some localization aspects has been provided and it is expected that the elementary structures here identified can be applied in the investigation of more enhanced Dirac-like systems, as for instance, in trapped ion \cite{diraclike01} and bilayer graphene platforms \cite{diraclike02} used to emulate relativistic Dirac equation properties. 

{\em Acknowledgments} -- This work was supported by the Brazilian agencies FAPESP (grant 2018/03960-9) and CNPq (grant 301000/2019-0).

\end{document}